\documentstyle[12pt,openbib,psfig]{article}
\def\be{\begin{equation}}
\def\ee{\end{equation}}

\begin{document}

\title{\bf Addendum to \\ ``Decoupling of pion coupling $f_{\pi}$ from quarks at
high density in three models, and its possible observational
consequences" \\
Phys. Lett. B 618 (2005) 115 \\ {\small DOI:
10.1016/j.physletb.2005.05.031} }\vskip .5cm
\author{Manjari Bagchi, Monika Sinha, Mira Dey, Jishnu Dey  } \vspace{.5 cm}
\date{\today }
\maketitle

In two recent papers the decoupling of the quarks from pions at
high density is totally ignored. In the more recent one - Baldo
$et ~al.$ \cite{baldo}, the authors {\bf assume} that the
$f_{\pi}$ remains constant and put the cut off as a variable with
increasing density. But from 1985 onward various groups including
Dey, Dey and Ghose \cite{ddg}, Dey and Dey \cite{dd}, Bernard,
Meissner and Zahed \cite{bmz} and Eletsky and Kogan \cite{ek} and
finally Bagchi $et~al.$ \cite{bsdd} have shown that in various
models including QCD sumrule, the pions decouple from quarks at
high density. The values of $f_{\pi}$ decreases with increasing
density. Baldo $et~al.$ considers  the model of Nambu and
Jona-Lasinio which is discussed in the work by Bagchi $et~al.$
\cite{bsdd}.

These papers \cite{ddg, dd, bmz,ek} also point out that the pion
coupling to the nucleon $g_{\pi NN}$ survives at high density.
This implies that the long range tensor force remains unaltered
at high density. Kl\"ahn $et ~al.$ \cite{16th} have recently
pointed out that constraints can be derived for the high-density
nuclear equation of state from phenomenology of compact stars and
heavy-ion collisions. Their scheme is applied to a set of
relativistic EoSs constrained otherwise from nuclear matter
saturation properties. The result is that no EoS can satisfy all
constraints, but those with density-dependent masses and coupling
constants appear most promising. Perhaps a density dependent form
for the tensor force will be very useful for deriving a new EoS.

Eletsky and Kogan \cite{ek} argued that it is worth asking a
question whether other hadronic couplings involving pions share
this feature. This may then affect hyperon stars which are
neutron stars with presence of strange baryons.

The second work is the more interesting one due to Soni and
Bhattacharya \cite{sb} where they suggest that the energy of ud
matter may be lowered in the presence of pion condensate. They
claim this imposes more stringent conditions for the stability of
uds matter which has to be more bound than ud matter - while ud
matter is less bound than nucleons in $Fe^{56}$. The conditions
should be revised with a density dependent $f_\pi$, and the
condition will slacken.

{}

\end{document}